\documentclass[12pt,times]{article}
\raggedright
\usepackage{parskip}
\usepackage{newtxtext, newtxmath}

\usepackage{geometry}
\usepackage{graphicx}
\usepackage{booktabs, multirow, array}
\usepackage[table]{xcolor}
\usepackage{pdfpages}
\usepackage{amsmath}
\usepackage{authblk}
\usepackage{rotating}
\usepackage{subcaption}
\usepackage{titlesec}

\titleformat*{\section}{\fontsize{14}{12}\bfseries}
\titleformat*{\subsection}{\fontsize{12}{12}\bfseries}
\titleformat*{\subsubsection}{\fontsize{12}{12}\bfseries}

\usepackage[font={footnotesize}]{caption}
\UseRawInputEncoding

\font\myfont=cmr12 at 16pt
\font\aufont=cmr12 at 12pt

\makeatletter
\renewcommand{\maketitle}{\bgroup\setlength{\parindent}{0pt}
\begin{flushleft}
  \myfont{\@title}
  
  \aufont{\@author}
\end{flushleft}\egroup
}
\makeatother

\title{\textbf{\fontsize{16}{12}{Utterance Clustering Using Stereo Audio Channels}}}

\author{Yingjun Dong$^{1, 2}$, Neil G. MacLaren$^{1, 3}$, Yiding Cao$^{1, 2}$, Francis J. Yammarino$^{1, 3}$, Shelley D. Dionne$^{1, 3}$, Michael D. Mumford$^{4}$, Shane Connelly$^{4}$, Hiroki Sayama$^{1, 2, 3}$, Gregory A. Ruark$^{5}$\\ 

$^{1}$Center for Collective Dynamics of Complex Systems, Binghamton University, State University of New York, Binghamton, NY 13902-6000, USA \\ 
$^{2}$Department of Systems Science and Industrial Engineering, Binghamton University, State University of New York, Binghamton, NY 13902-6000, USA \\
$^{3}$Bernard M. \& Ruth R. Bass Center for Leadership Studies, School of Management, Binghamton University, State University of New York, Binghamton, NY, USA \\
$^{4}$Department of Psychology, University of Oklahoma, Norman, OK, USA \\
$^{5}$U.S. Army Research Institute for the Behavioral and Social Sciences, Fort Belvoir, VA, USA \\
Correspondence should be addressed to Yingjun Dong; ydong25@binghamton.edu}
\date{}

\begin{document}

\maketitle

\section*{Abstract}
Utterance clustering is one of the actively researched topics in audio signal processing and machine learning. This study aims to improve the performance of utterance clustering by processing multichannel (stereo) audio signals. Processed audio signals were generated by combining left- and right-channel audio signals in a few different ways and then extracted embedded features (also called d-vectors) from those processed audio signals. This study applied the Gaussian mixture model for supervised utterance clustering. In the training phase, a parameter sharing Gaussian mixture model was conducted to train the model for each speaker. In the testing phase, the speaker with the maximum likelihood was selected as the detected speaker. Results of experiments with real audio recordings of multi-person discussion sessions showed that the proposed method that used multichannel audio signals achieved significantly better performance than a conventional method with mono audio signals in more complicated conditions.


\section*{Introduction}
With artificial intelligence (AI) development, many techniques are applied in our daily life, such as automatic speech recognition (ASR)~\cite{Menne_2019} and speaker recognition. Studies and products in speech processing are widely used in our daily life, such as Apple's Siri, Amazon's Alexa, Google Assistant, and Microsoft's Cortana. As more studies developed in speech processing, it will likely see further increases in popularity. Utterance clustering is a popular topic in speech processing that can be used for speaker diarization~\cite{anguera2012speaker} and ASR. However, most studies are based on laboratory datasets, and those cannot process the real-world problem very well. Both formal and informal meetings have more segments with overlapping speaking than segments with only one speaker~\cite{von2019all}. In the laboratory datasets, people speak one by one, but it is hard to ask people not to interrupt others' speech in the real world. The issue of overlapping speech segments has received considerable attention~\cite{chen2018multi}. To expand the application of speech processing, it is necessary to have better performance in overlapping utterance clustering.

 A key aspect of performance improvement in utterance clustering is audio feature embeddings. Feature embedding plays a vital role in ensuring the performance of utterance clustering. There are many studies that focus on the enhancement of audio feature embeddings, such as mel-frequency cepstral coefficient (MFCC)~\cite{davis1980comparison}, i-vector \cite{dehak2010front}, x-vector \cite{snyder2018x}, and d-vector \cite{variani2014deep}. However, a significant bottleneck towards the widespread adoption of speech processing applications in daily life is high-quality audio data requirements. Besides, audio signal processing could be a contributing factor to feature embeddings. There are needs for a better speaker diarization method using low-quality audio recording data in many social science experiments. This study initially tried several different published methods for our own experimental research, but their results were not as good as we had hoped.
 
 To address this problem, here a new method of audio signal processing was proposed for utterance clustering~\cite{dong2020speaker}. The challenge this study aims to address is how to handle low-quality audio data recorded in real-world discussion settings. The audio dataset was recorded using an ordinary video camcorder in a noisy environment without a professional microphone. This work contributes to the advance of utterance clustering when the recording conditions are limited.

This study aims to improve clustering performance by processing multichannel (stereo) audio signals. Mono audio signals are typically used in audio processing studies as they can be obtained easily by downmixing stereo audio signals. Then the d-vector of each audio segment was obtained using pre-trained neural networks as the audio feature representation. 

The Gaussian mixture model (GMM) was used as a supervised clustering method. The error rate (ER) of the clustering was compared, and the results showed that using the processed multichannel audio signal for utterance clustering was significantly better than using the original mono audio signal. The structure of the paper is as follows. Section 2 of this paper introduces some related work. In Section 3, the method in feature processing and the Gaussian mixture model are described. Section 4 shows the details of the dataset and the details of the experiments. The results are discussed in Section 5, conclusions and plans for future work are described in Section 6. 

\section*{Related Work}
Numerous researchers have made significant advances in utterance clustering and related fields over the last few decades. Certain studies place a greater emphasis on feature embeddings; historically, the most common feature representation was the MFCC~\cite{davis1980comparison}, which is a method based on the Fourier spectrum. Then, as factor analysis developed, Dehak et al.~\cite{dehak2010front} proposed a factor analysis called i-vector. Their factor analysis took into account the variability of speakers and channels without distinction. Lei et al.~\cite{lei_kun_2017} proposed wavelet packet entropy (WPE) to extract short vectors from utterances and then using the i-vector as a feature embedding. As with i-vectors, d-vectors~\cite{variani2014deep} also have fixed sizes regardless of the length of the input utterance. Wan et al.~\cite{wan2018generalized} trained speakers' utterances using a deep neural network, and the lengths of these utterances varied, resulting in fixed-length embeddings, namely d-vector. The distinction between i-vector and d-vector is that the former is generated using GMM, while the latter is trained using deep neural networks. Similar to the d-vector, the x-vector~\cite{snyder2018x} is also trained with deep neural networks. Ma et al.~\cite{ma_zuo_li_chen_2020} proposed an E-vector, which was obtained by minimizing the Euclidean metric to improve the performance of speaker identification. All of the feature embeddings described above are commonly utilized, and the d-vector was used in this study. There are also some works focus on the improvement of feature extraction. Lin et al.~\cite{lin_yumei_maosheng_defeng_chao_tonghan_2020} introduced a novel feature extraction approach that combines multiresolution analysis with chaotic feature extraction to improve the performance of utterance features. Daqrouq et al.~\cite{daqrouq_al-hmouz_balamash_alotaibi_noeth_2015} proposed a feature extraction method based on wavelet packet transform (WPT), they removed the silence parts from the audio data, then they decomposed the audio signal into wavelet packet tree nodes.

In some research, the clustering algorithms are given greater consideration. Delacourt et al.~\cite{delacourt2000distbic} applied the bayesian information criterion (BIC) to measure the distances among utterances, and then they conducted the agglomerative hierarchical clustering (AHC) based on the BIC metrics. Li et al.~\cite{li_yang_dai_2014} conducted GMM on MFCC to classify speakers' gender. Algabri et al.~\cite{algabri_mathkour_bencherif_alsulaiman_mekhtiche_2017} applied the Gaussian mixture model with the universal background model (GMM-UBM) to recognize speakers according to the MFCC of utterances. Shum et al.~\cite{shum2013unsupervised} used the re-segmentation algorithm of the Bayesian GMM clustering model based on i-vector to contribute to improving the speech clustering. Zajic et al.~\cite{zajic2017speaker} proposed a model for applying Convolutional Neural Network (CNN) on i-vector to detect speaker changes. Wang et al.~\cite{wang2018speaker} developed the LSTM model on d-vector for the speaker diarization. Zhang et al.~\cite{zhang2019fully} constructed a supervised speaker diarization system on the extracted d-vector, called unbounded interleaved-state recurrent neural networks (UIS-RNN).

In comparison to the previous efforts, this study used processed audio signals rather than mono audio samples. The processed audio signals are derived from multichannel (stereo) audio signals, and the proposed method attempted to preserve more representative audio characteristics.

\section*{Methods}

In this section, the proposed method of audio feature processing is discussed. The details of processing multichannel audio features are shown, and the tool which was employed to extract audio feature embeddings is described. Also, the clustering method is presented. 

\subsection*{Feature Processing}
\begin{sidewaysfigure}[p]
    \centering
    \includegraphics[width=\columnwidth]{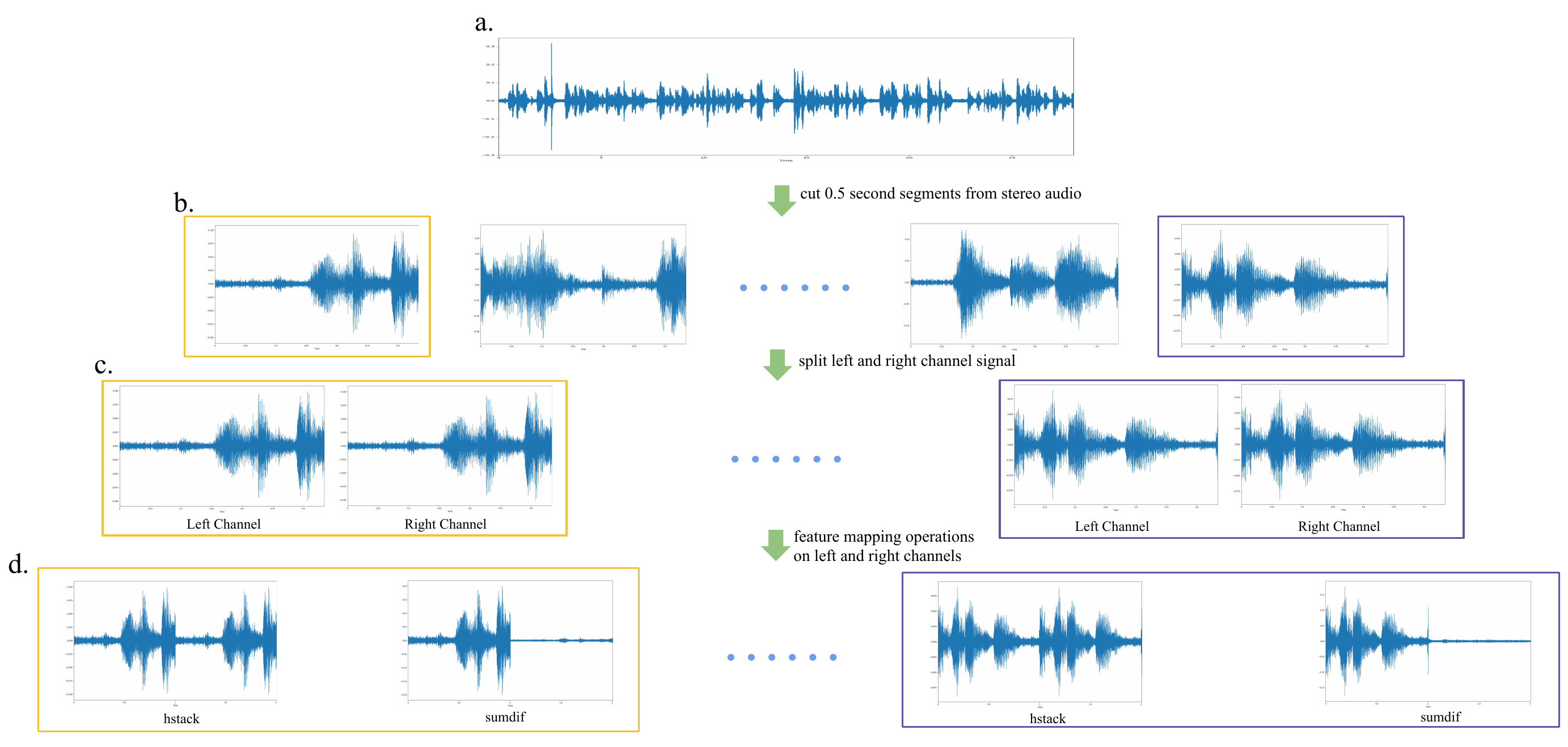}
    \caption{Visualization of audio signal processing for each speaker. The same color box represents the waveforms from the same speech segment. \textbf{a.} A stereo waveform of a speaker's speaking audio. \textbf{b.} Stereo waveforms in 0.5 second. \textbf{c.} Mono waveforms of extracted left and right channels audio for every 0.5 seconds. \textbf{d.} The processed waveforms for every 0.5 seconds.}
    \label{fig_sim}
\end{sidewaysfigure}

This study operated the left channel audio signals and the right channel audio signals to obtain speech-only audio features in the present work. 
The details of feature processing was visualized in Fig.~\ref {fig_sim}. This example shows that after removing the non-speech part, the speaker's speaking time is 27 seconds. In this work, 27 seconds of stereo audio were divided into 54 stereo audio segments, each of which is 0.5 seconds in length. After that, mono audio files were extracted, left channel audio files, and right channel audio files from the 0.5-second-long stereo audio files. The Python package librosa~\cite{brian_mcfee_2020_3606573} was used to obtain the left and right audio signals in the time series.

Horizontal stacking of the original left and right audio signals (hstack) and horizontal stacking of the sum and the difference of the left and right channel signals (sumdif) were performed. The computational complexity of the proposed method is still $O(L)$, where $L$ is the length of audio signal, which is the same in order as the traditional methods (although the actual computation takes about twice as much since our method processes two channels of audio signals).

For the training set, all speakers' utterances $S = (s_1,\, \ldots, \, s_i,\,\ldots,\, s_N)$ were acquired, where $N$ represents the number of speakers in the audio dataset, and $s_i$ represents the sequence of all speaking segments of the $i^{th}$ speaker. Specifically, $s_i = (x_{i, 1},\,\ldots,\, x_{i, t})$, where $x_{i, t}$ represents the $i^{th}$ speaker's audio signal at the $t^{th}$ segment. Then left and right channels audio from each segment in $s_i$ were extracted: $x_{i, t}^{L}$ for the $i^{th}$ speaker's left channel audio signal at the $t^{th}$ segment and $x_{i, t}^{R}$ for the $i^{th}$ speaker's right channel audio signal at the $t^{th}$ segment were obtained.
Using the left and right channels, the following two combined audio segments were created: $x_{i, t, \mathrm{hstack}} = (x_{i, t}^{L}, x_{i, t}^{R})$ and $x_{i, t, \mathrm{sumdif}} = (x_{i, t}^{L} + x_{i, t}^{R}, x_{i, t}^{L} - x_{i, t}^{R})$. For the $i^{th}$ speaker's all audio segments, $s_{i, W} = (x_{i, 1, W},\, \ldots,\, x_{i, t, W})$ was obtained, where $W \in \{\mathrm{hstack, sumdif}\}$. For a fair comparison, a mono stack was created, which is called mstack. It is a stack result of repeated mono signals, represented as $x_{i, t, \mathrm{mstack}} = (x_{i, t}^{mono}, x_{i, t}^{mono})$.

\subsection*{Feature Embeddings}
After feature processing, the d-vector~\cite{wan2018generalized} was extracted as the feature representation of the audio signals. The pre-trained model called Real Time Voice Cloning~\cite{realtimevoicecloning} was used to extract the d-vector. The pre-trained model was trained by using three datasets: one dataset is LibriSpeech ASR corpus \cite{panayotov2015librispeech} which contains 292,000 utterances for more than 2,000 speakers in English, and others are VoxCeleb 1 \& 2 \cite{Nagrani17}\cite{Chung18b} which contain over 1 million utterances for more than 7,000 speakers in multiple languages. 

A d-vector from each $s_{i, W}$ was extracted, to obtain $D_{i, W} = (d_{i, 1, W},\, \ldots,\, d_{i, t, W})$, where $D_{i, W}$ represents d-vectors of the $i^{th}$ speaker's all audio segments, and $W \in \{\mathrm{mono, mstack, hstack, sumdif}\}$. Then, GMM clustering on the extracted d-vectors was conducted. 

\subsection*{Gaussian Mixture Model}
The Gaussian mixture model (GMM) was used as the clustering method. GMM is one of the most frequently used tools for speakers clustering. In this study, separate GMM models for individual speakers were built, defined as:
\begin{align*} 
    p(y) = \sum^{M}_{m=1}\alpha_{m}\mathcal{N}(y;\mu_{m}, \Sigma_{m}),
\end{align*}
where $y$ represents the feature vector of audio signal, $p(y)$ is the probability that the input audio signal belongs to specific cluster, $\alpha_{m}$ represents the mixing proportions, $\mu_{m}$ represents mean, and $\Sigma_{m}$ represents covariance matrix~\cite{yu2016automatic}. The Expectation Maximization (EM) algorithm~\cite{bilmes1998gentle} was used to estimate the model parameters in GMM. GMM has significant advantages in acoustic modeling~\cite{yu2016automatic}. 

\section*{Experiments}

The details of the experiments are described in this section. The details of the dataset used in this study are presented firstly. Then, the tool used in this work to do audio processing is introduced. Also, the details of clustering experiments are shown. To ensure the comparison between the proposed method and comparative methods is fair, in the proposed experiments, the same audio data and the same audio processing method to extract multichannel audio signals and mono audio signal for both the proposed method and comparative methods were used. Last but not least, a parameter sharing GMM was conducted for both the proposed method and the comparative methods.

\subsection*{Dataset}
A dataset~\cite{MACLAREN2020101409} containing 11 video files of discussions by multiple participants in a real-world physical environment was used in the proposed work. One session with 4 members; 4 sessions with 6 members; 5 sessions with 7 members and one session with 8 members. The number of speakers in 11 videos ranged from 4 to 10, the number of female speakers varies from 1 to 6, the number of male speakers varies from 1 to 6, and all speakers speak English. Each speaker's speaking time ranged from 1 to 130.5 seconds. The total speaking time for all 11 videos is 31.6 minutes, and the average speaking time for each speaker is 26.7 seconds. The dataset was manually annotated with the ground-truth speaker labels.

In the proposed experiments, two comparison groups were set. Audio files in one group contain overlapping speeches, and in the other one, audio files do not contain overlapping speeches. These audio files in two groups are from the same audio. Speakers were in a real-world free discussion scenario and an ordinary video camcorder was used to record all videos and audios with a built-in stereo microphone.

\subsection*{Audio Processing}
FFmpeg~\cite{ffmpeg} was used to extract stereo audio files and mono audio files from video files. Based on the manually annotated speaking time data, audio segments for each of the different speakers were cut. Then, each audio segment of different speakers were cut into shorter segments of a length of 0.5 seconds. The audio files that were shorter than 0.5 seconds were deleted. Then, stereo signals into left and right channel signals were split, and d-vectors from processed signals were obtained. After the audio signal processing, the clustering experiments were conducted.  

\subsection*{Clustering by Gaussian Mixture Model}
The proposed work applied scikit-learn~\cite{scikit-learn} for GMM training and testing. In the initial experiment, a small part of the data set and traditional methods (mono audio signal) were used to adjust the parameters to obtain better accuracy. Then, for fair comparison, the same parameters were set for all the proposed methods. The full covariance type and K-means were used to initialize the model. 

The input for the clustering model is the d-vector, and clustering experiments were conducted 50 times; for each time, a 10-fold cross-validation test was conducted. 

In the training phase, there are d-vectors $D_{i, W}^{train} = (d_{i, 1, W}^{train},\, \ldots,\, d_{i, t_{1},W}^{train})$, where $t_1$ is $70\%$ of speaker $i$'s total speaking time. To train the model, the speakers' label sequence $Y_{i} = (y_{i, 1},\, \ldots,\, y_{i, t_{1}})$ for speaker $i$. This study trained GMM models for each speaker, then model set $M = (m_1,\, \ldots,\, m_N)$ was obtained, where $m_N$ represents the $N^{th}$ speaker's trained model. For the testing, there are d-vectors of audio segment for each speaker $D_{i, W}^{test} = (d_{i, 1, W}^{test},\,\ldots,\, d_{i, t_{2}, W}^{test})$, where $t_{2}$ is $30\%$ of speaker $i$'s total speaking time. For $N$ speakers, $D^{test} = (D_{1, W}^{test},\,\ldots,\, D_{N, W}^{test})$. Based on the model set $M$ and test set $D^{test}$, the results which had the maximum likelihood value to generate the prediction results $\hat{Y}$, then $\hat{Y}$ was compared with the ground truth $Y_{test}$ to obtain the error rate.

\section*{Results}
This part will show the results of proposed experiments. The visualization of feature vectors will be displayed to show the results of feature processing, then the results of GMM clustering and the results of significance tests will be shown. 
\subsection*{Feature Processing}
\begin{figure}[!h]
    \centering
    \includegraphics[width=\textwidth]{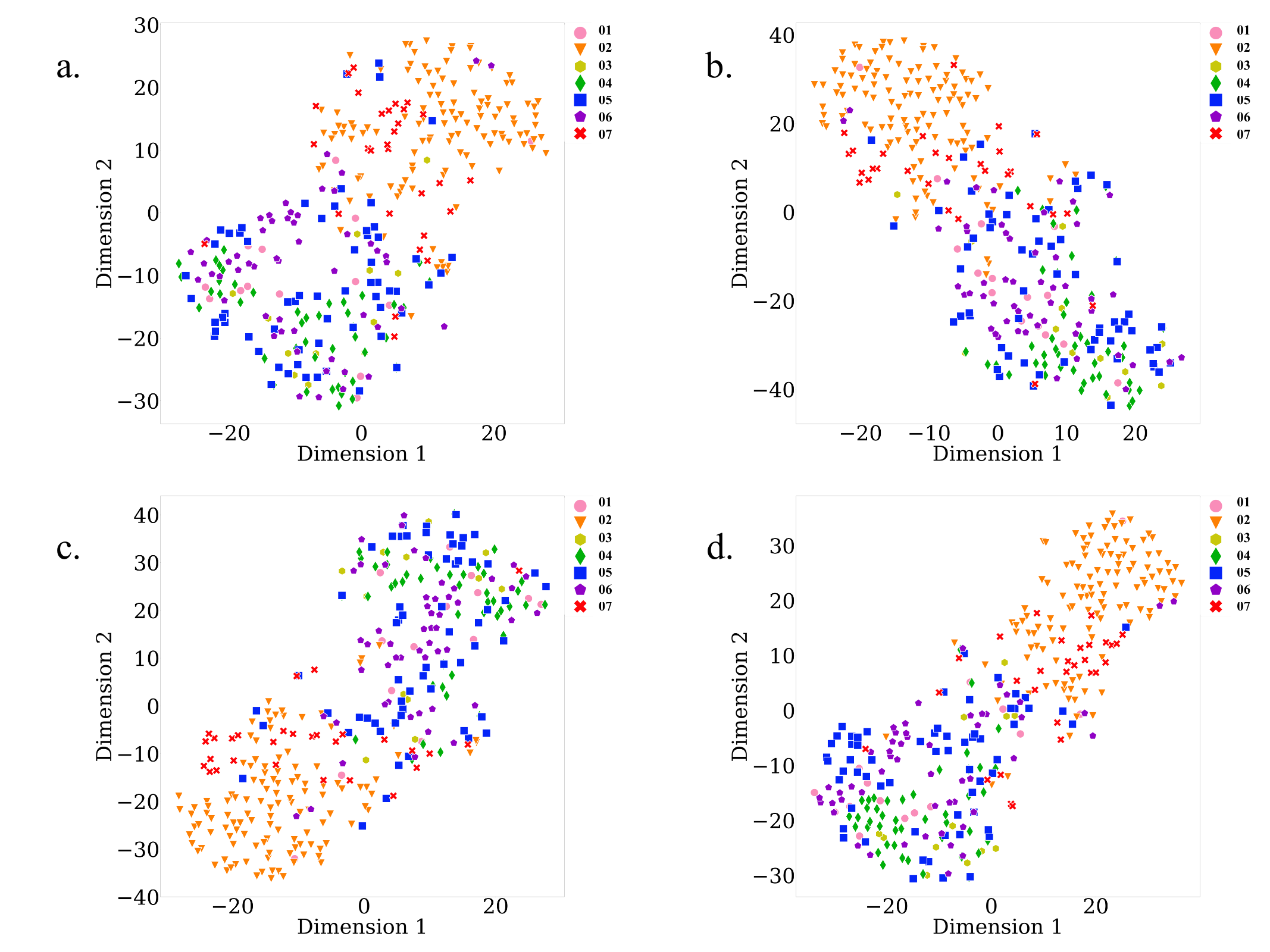}
    \caption{t-SNE visualization for seven speakers' feature vectors in the condition in which audio contains overlapping. Different colors represent different speakers. \textbf{a.} t-SNE visualization of d-vectors' clusters for speakers' mono signals. \textbf{b.} t-SNE visualization of d-vectors' clusters for speakers' mstack processed signals. \textbf{c.} t-SNE visualization of d-vectors' clusters for speakers' hstack processed signals. \textbf{d.} t-SNE visualization of d-vectors' clusters for speakers' sumdif processed signals.}
    \label{fig_dif_tsne}
\end{figure}

\begin{figure}[!h]
    \centering
    \includegraphics[width=\textwidth]{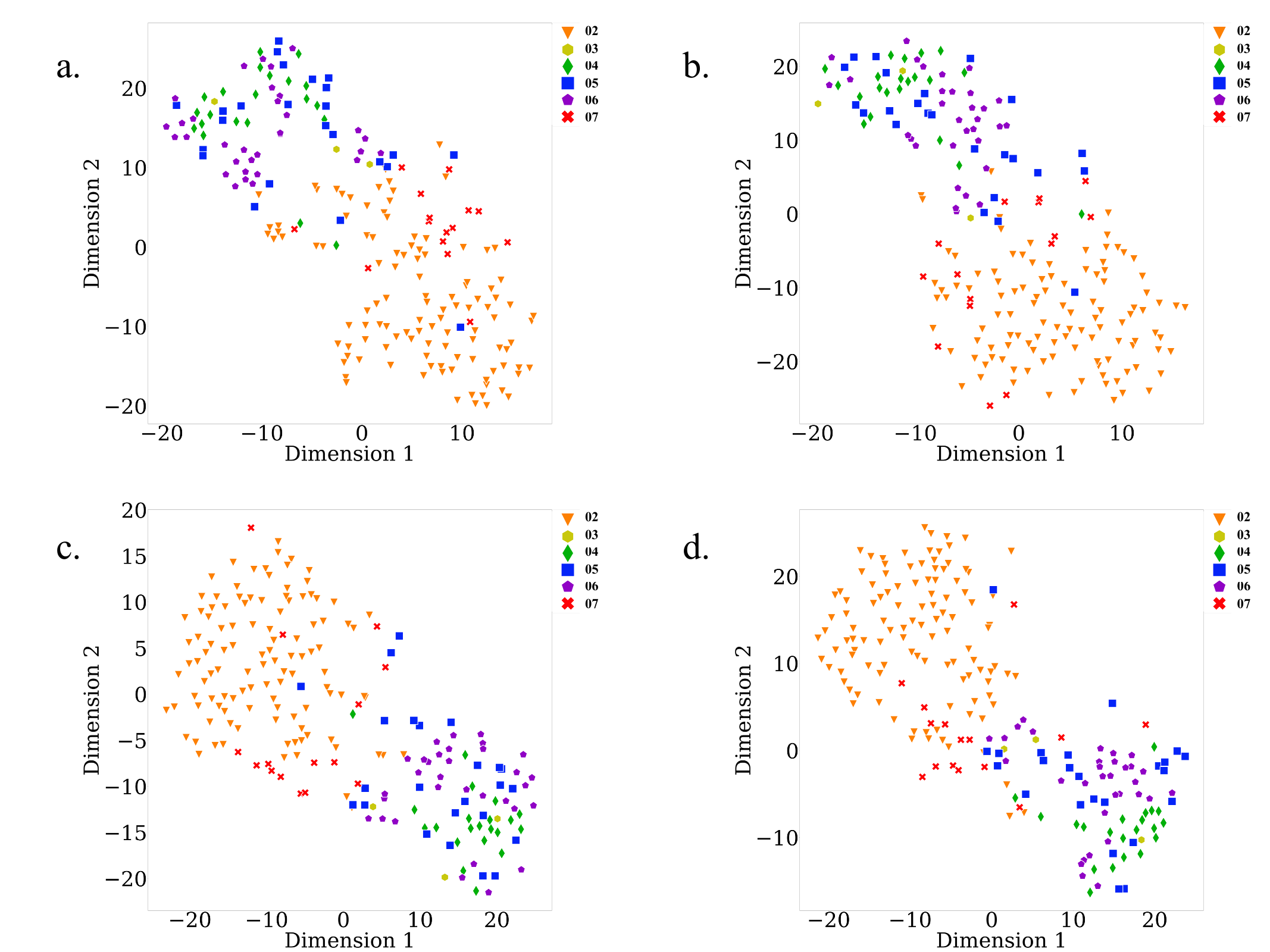}
    \caption{t-SNE visualization for seven speakers' feature vectors in the condition in which audio does not contain overlapping. Different colors represent different speakers. \textbf{a.} t-SNE visualization of d-vectors' clusters for speakers' mono signals. \textbf{b.} t-SNE visualization of d-vectors' clusters for speakers' mstack processed signals. \textbf{c.} t-SNE visualization of d-vectors' clusters for speakers' hstack processed signals. \textbf{d.} t-SNE visualization of d-vectors' clusters for speakers' sumdif processed signals.}
    \label{fig_dif_ntsne}
\end{figure}

Fig.~\ref{fig_dif_tsne} and \ref{fig_dif_ntsne} show the visualizations of processed feature vectors using t-SNE~\cite{maaten2008visualizing}, and the proposed algorithms (hstack and sumdif) show better clustering results. The data points show manifest clusters in the proposed methods. It can be seen from the Fig.~\ref{fig_dif_tsne} and Fig.~\ref{fig_dif_ntsne} that in the proposed methods, the data points of speaker 04, 05, 06 and 07 are clustered more closely. This implies that the d-vectors contain more information when the processed multichannel audio signal is used compared to the one extracted using the mono audio signal.

This study improves the performance of feature embeddings by processing multichannel audio signals. The proposed method extracts more useful features from the audio signals. Although the improvement is apparent, there are still some differences between the audio with overlap group and the audio without overlap group. Compared with the audio without overlap group, the proposed method enhances the performance of feature embedding in the audio with overlap group. The audio with overlap group is more intricate than the audio without overlap group. Extracting more useful features helps more in the complicated scenario than in the simple scenario.

\subsection*{Clustering}
\begin{table}[]
    \centering
    \caption{Mean z-scores of error rates on different methods}
    \begin{tabular}{ccccc}
         \midrule 
         Datasets & mono & mstack & hstack & sumdif  \\
         \hline
         with overlap & 0.2520 & 0.1861 & -0.1477 & \textbf{-0.2905} \\
         without overlap & 0.2764 & \textbf{-0.1419} & -0.0872 & -0.0473\\
         \midrule
    \end{tabular}
    \label{tab_zscore}
\end{table}

Table~\ref{tab_zscore} shows the comparison of the z-scores of GMM error rates in different algorithms. From Table~\ref{tab_zscore}, the sumdif algorithm works better than other algorithms in the audio with overlap group. In the audio without overlap group, the mstack algorithm works better. However, hstack and sumdif work better than the mono. The overall performance of hstack and sumdif is better than mono and mstack.

One-way ANOVA tests with Tukey HSD were performed to determine whether there were differences among the error rates of algorithms compared. Results are shown in Table~\ref{anova} for both the audio with overlap group and the audio without overlap group. Both groups had statistically very significant difference among the algorithms.

Results of the Tukey HSD test are shown in Table~\ref{tukey}. Results showed that the proposed algorithms (hstack, sumdif) are significantly different from traditional algorithms (mono, mstack) when the audio signals contained overlaps between speeches. The difference was less clear when the audio signals had no overlaps.

Results of clustering signify that even if the traditional GMM is applied instead of the deep learning model, using the processed audio signals in utterance clustering can achieve a higher accuracy score than mono audio signals. The dataset used in this study represents a real-world discussion setting. The proposed method shows significant improvements in a complicated discussion scenario, and the performance could be further improved by implementing deep learning models. The average of difference in means also shows that compared with simple condition (audio without overlap), the proposed method extracted more features from audio, which is more conducive to the utterance clustering in the complicated scenario (audio with overlap). 

\begin{table}[!h]
    \centering
    \caption{One-Way ANOVA Test Results}
    \begin{subtable}[c]{\textwidth}
        \centering
        \begin{tabular}{cccccc}
            \midrule
             & df & Sum of Squares & Mean Square & F & p \\
            \hline 
            Group & 3.0 & 2.2976 & 0.7659 & 49.1991 & 1.1082e-31 \\
            Residual & 21996.0 & 342.3983 & 0.0156 & & \\
            \midrule
        \end{tabular}
        \caption{with overlap}
        \label{over}
    \end{subtable}
    \quad
    \begin{subtable}[c]{\textwidth}
        \centering
        \begin{tabular}{cccccc}
            \midrule
             & df & Sum of Squares & Mean Square & F & p \\
            \hline 
            Group & 3.0 & 1.8214 & 0.6071 & 40.7833 & 2.8427e-26 \\
            Residual & 21996.0 & 327.4508 & 0.0149 & & \\
            \midrule
        \end{tabular}
        \caption{without overlap}
        \label{nonover}
    \end{subtable}
    \label{anova}
\end{table}

\begin{table}[!h]
    \centering
    \caption{Tukey HSD (FWER=0.05). Positive/negative meandiff means the average error rate of Group 2 is more/less than the average error rate of Group 1.}
    \begin{subtable}[c]{\textwidth}
        \centering
        \begin{tabular}{ccccccc}
            \midrule
            Group 1 & Group 2 & meandiff & p-adj & lower & upper & reject \\
            \hline 
            mono & mstcak & -0.0018 & 0.8611 & -0.0079 & 0.0043 & False\\
            mono & sumdif & -0.0234 & 0.001 & -0.0295 & -0.0173 & True \\
            mstack & sumdif & -0.0216 & 0.001 & -0.0277 & -0.0155 & True \\
            hstack & mono & 0.0187 & 0.001 & 0.0126 & 0.0248 & True \\
            hstack & mstack & 0.0169 & 0.001 & 0.0108 & 0.023 & True \\
            hstack & sumdif & -0.0047 & 0.1973 & -0.0108 & 0.0014 & False \\
            \midrule
        \end{tabular}
        \caption{with overlap}
        \label{hsd_over}
    \end{subtable}
    \quad
    \begin{subtable}[c]{\textwidth}
        \centering
        \begin{tabular}{ccccccc}
            \midrule
            Group 1 & Group 2 & meandiff & p-adj & lower & upper & reject \\
            \hline 
            mono & mstack & -0.0237 & 0.001 & -0.0297 & -0.0178 & True \\
            mono & sumdif & -0.0143 & 0.001 & -0.0203 & -0.0084 & True \\
            mstack & sumdif & 0.0094 & 0.001 & 0.0034 & 0.0154 & True \\
            hstack & mono & 0.0205 & 0.001 & 0.0146 & 0.0264 & True \\
            hstack & mstack & -0.0033 & 0.4938 & -0.0093 & 0.0027 & False \\
            hstack & sumdif & 0.0061 & 0.0425 & 0.0001 & 0.0121 & True \\
            \midrule
        \end{tabular}
        \caption{without overlap}
        \label{hsd_nonover}
    \end{subtable}
    
    \label{tukey}
\end{table}

\section*{Conclusions}
This study generated processed audio signals by combining left- and right-channel audio signals in two different ways. And then d-vectors were extracted as embedded features from those processed audio signals. The GMM was conducted for supervised utterance clustering. Based on the results obtained from the supervised clustering experiment, the proposed method works better in complicated conditions than traditional methods. Namely, the proposed method can achieve a higher accuracy score than using traditional algorithms in the speech that contains overlapping. This is because the stereo audio signals contain information about spatial location of the sound source (in a left-right direction space). In a typical real-world discussion setting, speakers tend to sit in a fixed location, so using spatial information can help speaker identification and utterance clustering. This study successfully demonstrated this idea. 

One limitation of the proposed method is the computational cost. Even though the theoretical computational complexity of the proposed method is the same as the traditional methods, in the actual experiments, the run time of our proposed method is greater than that of the traditional methods. Moreover, stereo audio signals were used in this study, so another limitation is that the input data must be multichannel audio signals that involve spatial information.  

In this study, GMM was applied as a clustering method. An innovative clustering model using deep learning will be proposed for future works. After applying different clustering methods, there are more comprehensive comparisons between the proposed algorithms and traditional algorithms. 

\section*{Data Availability Statement}
The audio data used to support the findings of this study have not been made available because Institutional Review Board permissions do not accommodate its release.

\section*{Conflicts of Interest}

The authors declare that there is no conflict of interest regarding the publication of this paper.

\section*{Funding Statement}

The research described herein was sponsored by the U.S. Army Research Institute for the Behavioral and Social Sciences, Department of the Army (Grant No. W911NF-17-1-0221).

The views expressed in this presentation are those of the author and do not reflect the official policy or position of the Department of the Army, DOD, or the U.S. Government.

\section*{Acknowledgments}
A preprint version of this work is also available from https://arxiv.org/abs/2009.05076.

\newpage
\bibliographystyle{IEEEtran}
\bibliography{ref}

\end{document}